# Influence of heavy metal materials on magnetic properties of Pt/Co/heavy metal tri-layered structures

Boyu Zhang, Anni Cao, Junfeng Qiao, Minghong Tang, Kaihua Cao, Xiaoxuan Zhao, Sylvain Eimer, Zhizhong Si, Na Lei, Zhaohao Wang, Xiaoyang Lin, Zongzhi Zhang, Mingzhong Wu, and Weisheng Zhao

# Influence of heavy metal materials on magnetic properties of Pt/Co/heavy metal tri-layered structures

Boyu Zhang,[1,2,3,a)] Anni Cao,[1,2,3,a)] Junfeng Qiao,[1,2,3] Minghong Tang,[4] Kaihua Cao,[1,2,3] Xiaoxuan Zhao,[1,2,3] Sylvain Eimer,[1,2,3] Zhizhong Si,[1,2,3] Na Lei,[1,2,3] Zhaohao Wang,[1,2,3] Xiaoyang Lin,[1,2,3,b)] Zongzhi Zhang,[4] Mingzhong Wu,[5] and Weisheng Zhao[1,2,3,b)]

[1]Fert Beijing Research Institute, Beihang University, Beijing 100191, China
[2]Beijing Advanced Innovation Center for Big Data and Brain Computing (BDBC), Beihang University, Beijing 100191, China
[3]School of Electronic and Information Engineering, Beihang University, Beijing 100191, China
[4]Key Laboratory of Micro and Nano Photonic Structures (Ministry of Education), Department of Optical Science and Engineering, Fudan University, Shanghai 200433, China
[5]Department of Physics, Colorado State University, Fort Collins, Colorado 80523, USA

Pt/Co/heavy metal (HM) tri-layered structures with interfacial perpendicular magnetic anisotropy (PMA) are currently under intensive research for several emerging spintronic effects, such as spin-orbit torque, domain wall motion, and room temperature skyrmions. HM materials are used as capping layers to generate the structural asymmetry and enhance the interfacial effects. For instance, the Pt/Co/Ta structure attracts a lot of attention as it may exhibit large Dzyaloshinskii-Moriya interaction. However, the dependence of magnetic properties on different capping materials has not been systematically investigated. In this paper, we experimentally show the interfacial PMA and damping constant for Pt/Co/HM tri-layered structures through time-resolved magneto-optical Kerr effect measurements as well as magnetometry measurements, where the capping HM materials are W, Ta, and Pd. We found that the Co/HM interface play an important role on the magnetic properties. In particular, the magnetic multilayers with a W capping layer features the lowest effective damping value, which may be attributed to the different spin-orbit coupling and interfacial hybridization between Co and HM materials. Our findings allow a deep understanding of the Pt/Co/HM tri-layered structures. Such structures could lead to a better era of data storage and processing devices. *Published by AIP Publishing.* [http://dx.doi.org/10.1063/1.4973477]

Recently, perpendicularly magnetized materials have attracted significant interest owing to their high anisotropy, low switching current, and high scalability. These features could enable a leading class of memory and logic devices.[1–3] Pt in contact with Co is well known to generate an interfacial perpendicular magnetic anisotropy (PMA) with a (111) texture,[4–9] and the Pt/Co/heavy metal (HM) tri-layered structures are under intense investigation to explore a number of emerging spin-related effects, such as spin-orbit torque (SOT),[10–14] domain wall motion,[15,16] and room temperature skyrmions.[17–19] For instance, Co/Pt-based multilayers with large PMA were applied as bottom pinned layers in perpendicular magnetic tunnel junctions.[9] Pt/Co/Ta and Pt/Co/Ir structures were found to exhibit strong Dzyaloshinskii-Moriya interaction (DMI) for their asymmetric stack structures.[18,19] In addition to the interfacial PMA, the damping constant is an important magnetic parameter in Pt/Co/HM tri-layered structures as it determines the magnetization dynamics, such as the speed of magnetization reversal[20] and domain-wall motion.[21,22] It also affects the thermal fluctuations and noise levels for magnetic read head and sensor applications.[23] A series of studies have focused on the physical origin of the damping constant[24–28] as well as its correlation with PMA,[29,30] and the damping constant was mostly investigated by the thickness variation in the Co thin film structures with a strong PMA.[31–33] However, the dependence of magnetic properties on different capping materials, especially HM materials, has not been systematically studied. In this work, we will investigate experimentally the interfacial PMA and damping constant in Pt/Co/HM tri-layered structures, and analyze the influence of capping HM materials on magnetic properties. The origins of magnetic properties are elucidated by examining the physical contributions of different capping materials. Our findings will provide helpful information for the design of magnetic multilayers with desired magnetic properties, and may lead to a better era of data storage and processing devices.

The samples were grown by sputtering. They are stacks of Pt/Co/HM composed of a 0.8-nm-thick Co layer sandwiched between a 3-nm-thick Pt layer and a 2-nm-thick HM capping layer, where the capping HM materials are W, Ta, or Pd. The stacks were prepared using a Ta (2 nm) seed layer, which ensured the (111) texture of Pt, while the top Pt (3 nm) film formed a protective layer preventing the oxidation of the films, as shown in Fig. 1(a). Thin film magnetic characteristics were studied by using alternating gradient field magnetometers (AGFM) at room temperature to obtain the interfacial PMA index $K_{eff}*t_{eff}$, with the effective magnetic anisotropy energy $K_{eff}$ as follows:[34]

---

[a)]B. Zhang and A. Cao contributed equally to this work.
[b)]Authors to whom correspondence should be addressed. Electronic addresses: XYLin@buaa.edu.cn and weisheng.zhao@buaa.edu.cn

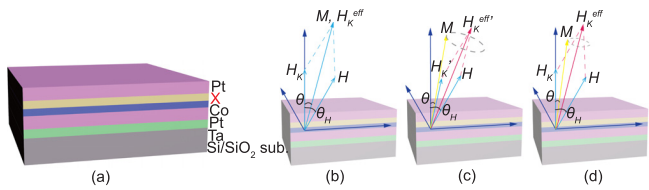

FIG. 1. (a) Schematic Ta/Pt/Co/Capping layer/Pt stack structure and (b)–(d) damping precession process.

$$K_{eff} = \frac{1}{2} H_K M_S, \quad (1)$$

where $H_K$ and $M_S$ represent the anisotropy field and the saturation magnetization, respectively. The interfacial PMA constant $K_i$ is determined by the relation[34,35]

$$K_{eff} = \frac{K_i}{t_{eff}} + (K_V - 2\pi M_S^2), \quad (2)$$

where $K_v$ is the bulk anisotropy and the term of $-2\pi M_S^2$ represents the demagnetizing energy of the shape anisotropy.

The damping constant is deduced from field-dependent time-resolved magneto-optical Kerr effect (TR-MOKE) measurements.[31,36,37] The sketch map in Fig. 1(b) shows a modification of the effective field $H_K^{eff}$ as a result of the applied field $H$ when the sample is probed prior to excitation. Then, the fast demagnetization process occurs due to the pulsed laser excitation. The magnitude, the direction of the magnetization $M$, and the anisotropy field $H_K$ change since the lattice is changed by the laser heat, thereby altering the equilibrium orientation. $M$ starts to precess around its reestablished equilibrium as the electronic thermal bath equilibrates with lattice (Fig. 1(c)). After the heat diffuses away, a slower relaxation follows and $H_K$ is restored, but the precession continues because of the initial displacement of $M$ (Fig. 1(d)). These relaxation processes are related to the specific heats and the coupling between different energy systems. In our TR-MOKE measurements, the beam wavelength and the pump beam fluence were set to 800 nm and 4 mJ/cm$^2$, respectively. The probe beam, whose intensity is much less than that of the pump beam, was almost normally incident on the film surface. TR-MOKE measurements were obtained with an applied field $H$, varying from 5 kOe to 15 kOe. The angle $\theta_H$ between the applied field and the film normal direction was set to 71°.

Fig. 2 shows the hysteresis loops of Ta (2 nm)/Pt (3 nm)/Co (0.8 nm)/Capping layer (2 nm)/Pt (3 nm) stacks with the magnetic fields in-plane or out-of-plane to the stacks, where the capping materials are Ta, W, and Pd. With the saturation field $H_S$ smaller than 200 Oe from the out-of-plane curve, the samples of Pt/Co/Ta and Pt/Co/Pd show the perpendicular magnetization. The Co thickness dependence of $M_S$ indicates the existence of a magnetic dead layer (thickness $t_{dead}$) when the capping material atoms diffuse into the Co layer during the sputtering deposition process. The effective Co thickness $t_{eff}$ is estimated from the magnetization as a function of the Co thickness, which is shown in Table I. The interfacial PMA index $K_{eff}*t_{eff}$, derived from the saturation magnetization $M_S*t_{eff}$ and the anisotropy field $H_K$, could be estimated by using Eq. (1). The thin film structure with a Ta capping layer has a $K_{eff}*t_{eff}$ value of 0.28 erg/cm$^2$, which is much larger than 0.09 erg/cm$^2$ for Pd and 0.05 erg/cm$^2$ for W, indicating a strong PMA of Pt/Co/Ta tri-layers. The interfacial PMA constant $K_i$, obtained by varying $t_{eff}$ and the corresponding $K_{eff}*t_{eff}$ under Eq. (2), shows also a relatively large value for Ta-capped stack. Apart from the contribution of the same bottom Pt/Co interface, the Co/Ta interface shows a larger interfacial PMA contribution than Co/Pd and Co/W interfaces. An example of dead layer and interfacial PMA constant calculation for Pt/Co/Pd tri-layered structures is described in Part I of the supplementary material. By comparing the hysteresis loops of Pt/Co/HM tri-layered structures with different Co thickness, we find that the spin-reorientation transition thickness from out-of-plane to in-plane is from 0.65 nm to 0.71 nm, as shown in Fig. S1 of the supplementary material.

The time-domain TR-MOKE signals were fitted by the following equation:[38]

$$\theta_k = a + b e^{-\frac{t}{\tau_0}} + c \sin(2\pi f t + \varphi) e^{-\frac{t}{\tau}}, \quad (3)$$

where $a + b e^{-\frac{t}{\tau_0}}$ represents an exponential decay background, $c$ and $f$ are the amplitude and the frequency of the magnetization precession, respectively, $\varphi$ donates the initial phase of the oscillation, and $\tau$ stands for the relaxation time related to the field-dependent damping $\alpha$ by the relation $\alpha = 1/(2\pi f \tau)$. The best fits to the experimental data of the Pt/Co/Pd thin film structure are shown in Fig. 3(a). Fig. 3(b) exhibits a monotonic decrease in the precession frequency $f$ with decreasing $H$, and the data are obtained from the fast Fourier transform (FFT) of the time-domain signals. Fig. 4 presents the field-dependent damping $\alpha$ of each sample with the applied field values varying from 5 kOe to 15 kOe. It can be estimated that 0.8 nm Co contains less than four layers of Co atoms, indicating that the abnormal points may be attributed to the non-uniformity of multilayers. In comparison with the Pd-capped sample, the W-capped sample has a lower damping value, around 0.03–0.04, while the Ta-capped sample shows a higher damping value, 0.06–0.08. It is important to mention that these field-dependent damping values include extrinsic contributions from two-magnon scattering (TMS)[40] and inhomogeneous line broadening (ILB).[41] They are both associated with the spatial inhomogeneity of the properties of the thin-film samples.

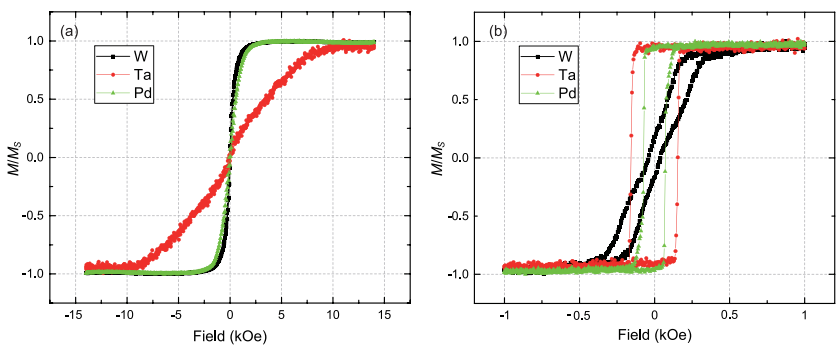

FIG. 2. Hysteresis loops with the field in-plane (a) and out-of-plane (b) for Ta (2 nm)/Pt (3 nm)/Co (0.8 nm)/capping layer (2 nm)/Pt (3 nm) stacks with capping HM materials W, Ta, and Pd.

TABLE I. Experimental magnetic properties of Ta (2 nm)/Pt (3 nm)/Co (0.8 nm)/capping layer (2 nm)/Pt (3 nm) thin film structures.

| Capping materials | $H_c$ (Oe) | $M_S*t_{eff}$[a] (emu/cm$^2$) | $H_K$[b] (Oe) | $t_{dead}$ (nm) | $t_{eff}$[c] (nm) | $M_S$ (emu/cm$^3$) | $K_{eff}*t_{eff}$ (erg/cm$^2$) | $K_i$ (erg/cm$^2$) | $K_{eff}$ (erg/cm$^3$) | $\alpha_{eff}$ |
|---|---|---|---|---|---|---|---|---|---|---|
| W  | 38  | $0.91 \times 10^{-5}$ | 1120 | 0.37 | 0.43 | 2090 | 0.05 | 0.52 | $1.2 \times 10^6$ | 0.033 |
| Ta | 155 | $0.76 \times 10^{-5}$ | 8100 | 0.43 | 0.37 | 2034 | 0.28 | 0.59 | $7.4 \times 10^6$ | 0.063 |
| Pd | 72  | $1.23 \times 10^{-5}$ | 1520 | 0.16 | 0.64 | 1926 | 0.09 | 0.55 | $1.5 \times 10^6$ | 0.054 |

[a]$M_S*t_{eff}$ is obtained by dividing the measured moment by the area of the Co magnetic layer.
[b]$H_K$ is obtained by extracting the field corresponding to 90% of measured moment in the hard axis.
[c]$t_{eff}$ is obtained by subtract $t_{dead}$ from the Co thickness $t$.

Based on these field-dependent damping results, the effective damping $\alpha_{eff}$ can be deduced from the equations as follows:[33,39]

$$\tau^{-1} = |\gamma|\alpha_{eff}\frac{H_1 + H_2}{2}, \quad (4a)$$

$$H_1 = H\cos(\theta_H - \theta) + H_K^{eff}\cos 2\theta, \quad (4b)$$

$$H_2 = H\cos(\theta_H - \theta) + H_K^{eff}\cos^2\theta, \quad (4c)$$

where $|\gamma|$ is the absolute gyromagnetic ratio from the TR-MOKE fitting results, $\theta_H$ is the angle of the applied field H related to the film normal direction, and $\theta$ is the angle between the magnetization vector and the film normal direction.

Based on the field-dependent experimental data, $\alpha_{eff}$ could be obtained as 0.033, 0.063, and 0.054 for W, Ta, and Pd, respectively (Table I). Although it has been reported in other material systems that Pd and Pt contribute more to the damping in contrast with Ta,[24,42] Ta-capped stack presents a relatively large $\alpha_{eff}$ compared with Pd and W in our Pt/Co/HM tri-layered structure, which we will discuss in the following section. The existence of spin pumping and intermixing contribution implies that the $\alpha_{eff}$ presents an upper limit of the intrinsic damping.[33,43,44] However, contributions of spin pumping effect and intermixing effect can be evaluated to be not dominant in our samples. (Details can be seen in Part IV of the supplementary material). In magnetic multilayer samples, there can also be some extrinsic contributions because of TMS[40] and ILB,[41] which could be the main origin of the field-dependent feature shown in Fig. 4. However, TMS and ILB are relatively small in our samples for the following reasons. First, the TMS contribution usually shows a peak response in the field or frequency dependent damping results, which originates from the dependence of the spin-wave manifold on the magnetic field. The absence of obvious peaks in our results (Fig. 4) thus suggests that the TMS contribution is relatively small. Second, the ILB contribution to the field-dependent damping usually decreases as the magnetic field increases.[45] The effective damping values presented in our work are close to the high-field data (15 kOe, see Fig. 4), so the ILB contribution should be relatively weak.

Table I shows a dependence of the effective damping $\alpha_{eff}$, the interfacial PMA index $K_{eff}*t_{eff}$, and the interfacial PMA constant $K_i$ on the capping materials. We will perform further theoretical analysis for the influence of capping HM materials on magnetic properties. The interfacial PMA originates from the adjustment of orbital momentum due to the orbital hybridization at interfaces. For instance, the hybridization of Co-3d and Pt-5d orbitals at Co/Pt interface induces the PMA. The variation of the interfacial PMA with different capping materials may be attributed to the different orbital hybridizations via spin-orbit coupling, which is between the magnetic field created by electron's orbital motion around the nucleus and its spin. For our uniaxial thin film structures, an estimate of the effective magnetic anisotropy energy $K_{eff}$ is[46]

$$K_{eff} \sim \frac{\xi^2}{W}, \quad (5)$$

where $\xi$ is the spin-orbit constant and $W$ is the $d$ bandwidth. We consider $\xi$ as the spin-orbit coupling influence of Pt and capping materials on Co at the two interfaces of Pt/Co/HM stacks[47] and $W$ as the spread of density of states (DOS) projected onto the $d$ orbital in this structure from the first-principles calculations based on the Vienna ab initio simulation package (VASP).[48,49] The first-principles calculation results can be found in Fig. 5 and Table II. Both the calculated $\xi^2/W$ and the experimental results $K_{eff}$ show the largest interfacial PMA in the sample with Ta capping layer and the smallest in the W-capped stack (Tables I and II). The influence of HM materials on the interfacial PMA may be explained by the different spin-orbit coupling and the hybridization of Co/HM interface, which change the spin-orbit constant[46,47] and the $d$ bandwidth.[29,31] The intrinsic damping $\alpha_{in}$ can be characterized as[50]

$$\frac{1}{|\gamma|M_S}\mu_B^2 D(E_F)\frac{(g-2)^2}{\tau}. \quad (6)$$

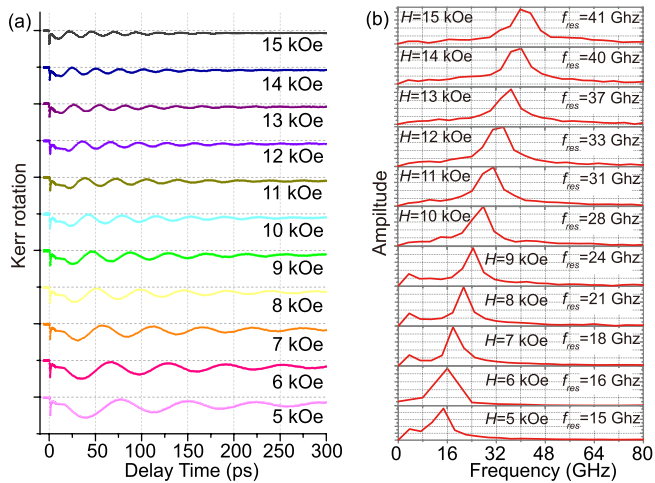

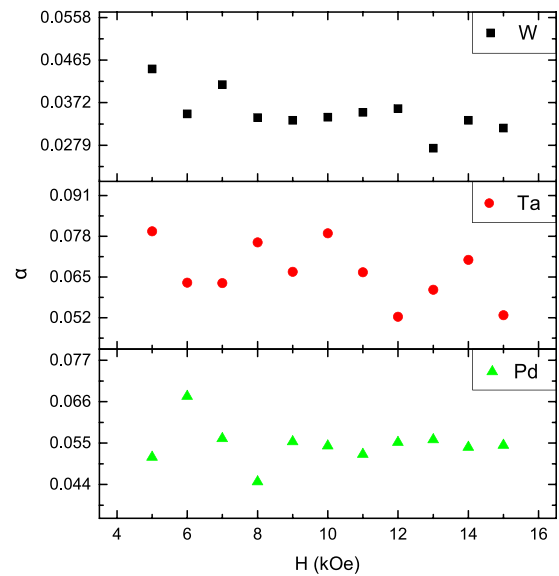

FIG. 3. (a) Time-resolved Kerr signals of Pt/Co/Pd thin film structure at different applied fields. (b) The corresponding FFT spectra, where the precession frequencies corresponding to each field value are indicated.

FIG. 4. The field-dependent damping $\alpha$ as a function of the applied field for Ta (2 nm)/Pt (3 nm)/Co (0.8 nm)/capping layer (2 nm)/Pt (3 nm) stacks with capping HM materials W, Ta, and Pd.

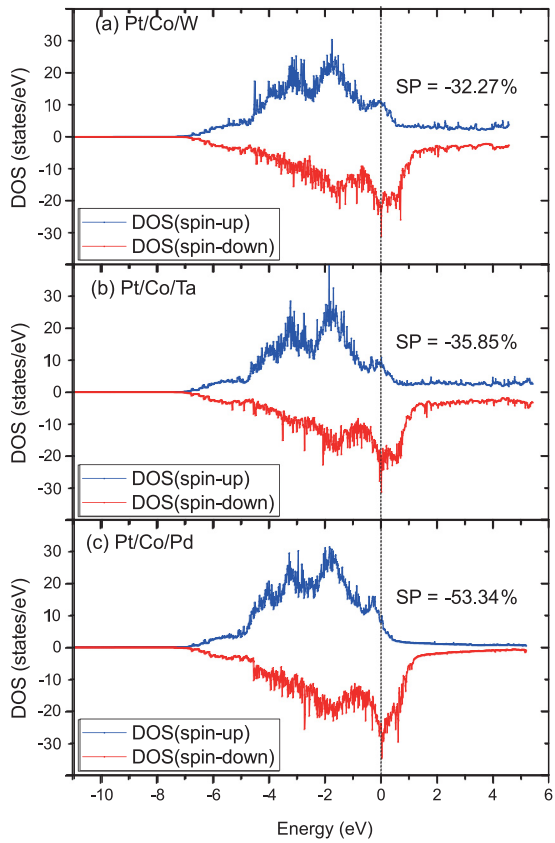

FIG. 5. Majority-spin (positive) and minority-spin (negative) DOS on the $d$ orbital in the (a) Pt/Co/W, (b) Pt/Co/Ta, and (c) Pt/Co/Pd thin film structures, where SP is the spin polarization at the Fermi energy (which is set to zero).

As the last term $(g-2)^2/\tau$ is proportional to $\xi^2/W$,[51] the following equation is obtained:

$$\alpha_{in} \sim \frac{1}{|\gamma|M_S}\mu_B^2 D(E_F)\frac{\xi^2}{W}, \quad (7)$$

where $(g-2)^2$ is the deviation of the $g$ factor from the free-electron value, $1/\tau$ is the ordinary electron orbital scattering frequency, $D(E_F)$ is the total DOS at the Fermi energy projected onto the $d$ orbital in the Pt/Co/HM stacks according to the first-principles calculations (see Fig. 5 and Table II) and $\mu_B$ is the Bohr magneton. As a main influence on the intrinsic damping, the magnon-electron scattering (MES) facilitates an energy transfer from magnetic sub-systems to non-magnetic sub-systems, resulting in a capping-material-dependent contribution to $\alpha_{eff}$. Many theories have been put forth to explain damping as the result of MES, such as the s-d exchange model,[52] the breathing Fermi surface (BFS) model,[53] and Kamberský's torque correlation model (TCM).[51] MES depends on not only the spin-orbit coupling which relates to $\xi$, but also can be influenced by the hybridization of Co/HM interface which changes $D(E_F)$ and $W$.[51–53] In general, the as-measured damping can include the contribution of intrinsic damping and those from the spin pumping effect and interfacial intermixing.[43,44] The as-calculated results (Table II), which provide theoretical estimations of the intrinsic damping, suggest a lowest damping value for magnetic multilayers with a W capping layer and follow the same trend of the as-measured effective damping. Our further experiment results indicate that (1) spin pumping plays a minor role in our samples; (2) the contribution of the intermixing effect will be capping material dependent, but such a damping enhancement should be slight (see Part IV of the supplementary material). As a result, the intrinsic damping dominates the relaxation in our samples. In addition, although the $\alpha_{eff}$ from experimental measurement may contain slight contributions from spin pumping and intermixing, our results are significant and realistic to the application based on Pt/Co/HM tri-layered structures considering the fact that both contributions exist in the magnetic multilayers.

In summary, we experimentally present the magnetic property dependence on the capping HM materials of Pt/Co/HM tri-layered structures. Three HM materials have been investigated through measurements performed by the magnetometry and all-optical pump-probe techniques. The damping constant and interfacial PMA are both sensitive to the Co/HM interface. The magnetic multilayers with a W capping layer features the lowest effective damping value, which may be attributed to the different spin-orbit coupling and interfacial hybridization between Co and HM materials. Our findings suggest that the use of PMA thin films with different capping HM materials can offer another degree of freedom for spintronic memory and logic device design.

TABLE II. Calculations of $K_{eff}$ and $\alpha_{in}$ of Ta (2 nm)/Pt (3 nm)/Co (0.8 nm)/capping layer (2 nm)/Pt (3 nm) thin film structures.

| Capping materials | $\xi$ (eV) | $W$ (eV) | $\xi^2/W$ (meV) | $|\gamma|$ (Grad/s Oe) | $M_S$ (emu/cm$^3$) | $D(E_F)$ (states/eV) | $D(E_F)\xi^2 / |\gamma|M_S W$ |
|---|---|---|---|---|---|---|---|
| W | 0.95 | 18.5 | 49 | 0.0191 | 2090 | 32.6 | 0.04 |
| Ta | 0.90 | 11 | 73.3 | 0.0180 | 2034 | 28.9 | 0.065 |
| Pd | 0.79 | 12 | 51.9 | 0.0185 | 1926 | 34.1 | 0.052 |

See supplementary material for: I. Calculation of dead layer and interfacial PMA constant for Pt/Co/Pd tri-layered structures. II. Spin-reorientation transition thickness. III. Precession frequency as a function of the applied field from TR-MOKE signals and the fit to the Kittel equation. IV. Contributions of the spin pumping effect and interfacial intermixing on the damping.

The authors would like to thank the support from the projects from National Natural Science Foundation of China (Nos. 61571023, 61501013, 61471015, 61627813, and 51602013), Beijing Municipal of Science and Technology (No. D15110300320000) and the International Collaboration Project from the Ministry of Science and Technology in China (No. 2015DFE12880), the Program of Introducing Talents of Discipline to Universities in China (No. B16001). The authors also thank the fruitful discussions with Professor Albert Fert.